\documentclass[preprint,12pt]{aastex}           
\usepackage{natbib,graphics}

\newcommand{\be}{\begin{equation}}
\newcommand{\ee}{\end{equation}}

\newcommand{\bea}{\begin{eqnarray}}
\newcommand{\eea}{\end{eqnarray}}

\shorttitle{QPOs from Diskoseismology}
\shortauthors{Wagoner, Silbergleit \& Ortega-Rodr\'{\i}guez}

\begin{document}
\title{`Stable' QPOs and Black Hole Properties from Diskoseismology} 

\author{Robert V. Wagoner\altaffilmark{1}} 
\affil{Department of Physics and Center for Space Science and Astrophysics \\ 
Stanford University, Stanford, CA 94305--4060}
\author{Alexander S. Silbergleit\altaffilmark{2}}
\affil{Gravity Probe B, Stanford University, Stanford, CA 94305--4085}
\and\author{Manuel Ortega-Rodr\'{\i}guez\altaffilmark{3}}
\affil{Department of Applied Physics and Gravity Probe B, \\
Stanford University, Stanford, CA 94305--4090}

\altaffiltext{1}{wagoner@stanford.edu} 
\altaffiltext{2}{gleit@relgyro.stanford.edu} 
\altaffiltext{3}{manuel@stanford.edu}

\begin{abstract}

We compare our calculations of the frequencies of the fundamental g, c, and p--modes of relativistic thin accretion disks with recent observations of high frequency QPOs in X-ray binaries with black hole candidates. These classes of modes encompass all adiabatic perturbations of such disks. The frequencies of these modes depend mainly on only the mass and angular momentum of the black hole; their weak dependence on disk luminosity is also explicitly indicated. Identifying the recently discovered relatively stable QPO pairs with the fundamental g and c modes provides a determination of the mass and angular momentum of the black hole. For GRO J1655--40, $M=5.9\pm 1.0 M_\sun$, $J=(0.917\pm 0.024)GM^2/c$, in agreement with spectroscopic mass determinations. For GRS 1915+105, $M=42.4\pm 7.0 M_\sun$, $J=(0.926\pm 0.020)GM^2/c$ or (less favored) $M=18.2\pm 3.1 M_\sun$, $J=(0.701\pm 0.043)GM^2/c$. We briefly address the issues of the amplitude, frequency width, and energy dependence of these QPOs.

\end{abstract}


\section{Introduction}

A few years ago, we predicted that the angular momentum of the candidate black hole in the X-ray binary GRO J1655--40 is approximately 93\% of maximum \citep{w}. (The dimensionless angular momentum parameter $a=cJ/GM^2$ is less than unity in absolute value.) This was based upon the identification of the `stable' 300 Hz feature with the fundamental g--mode in its accretion disk \citep{per}, plus the mass determination of $7.02\pm 0.22\; M_\sun$ from observations of its companion by \citet{OB}. As we shall see below, the fundamental c--mode should then be at a frequency near 450 Hz, in agreement with the recent discovery of another apparently stable feature in its power spectrum by \citet{S1655}. A second relatively stable high frequency QPO has also been recently seen in another galactic microquasar, GRS 1915+105 \citep{S1915}.

For the past ten years, our group has investigated the normal modes of oscillation of standard (geometrically thin) equilibrium models of black hole accretion disks. The early work of various groups has been reviewed by \citet{kfm} and \citet{w}. We perturb the optically thick fully relativistic models of \citet{nt,pt}. We have usually considered accretion disks which are barotropic [$p=p(\rho)$], producing a vanishing buoyancy frequency. The results below are relatively insensitive to the value of the viscosity parameter, here taken to be $\alpha = 0.1$. 

The key frequencies, associated with free-particle orbits in the disk, are the rotational [$\Omega(a,r)$], vertical epicyclic [$\Omega_\perp(a,r)$], and radial epicyclic [$\kappa(a,r)$] angular frequencies, respectively.  
The effective inner edge of the disk is close to the radius $r_i(a)$ of the last stable circular orbit, where $\kappa(a,r_i)=0$. For $a>0$, which we shall henceforth assume, $\Omega(r)>\Omega_\perp(r)>\kappa(r)$. 

We stress that linear combinations of these modes should describe all adiabatic perturbations of such disks. There are essentially three classes of modes, designated g, c, and p. We shall focus mainly on the fundamental mode within each class, corresponding to the minimum number of nodes in the axial, radial, and vertical components of the eigenfunction $\delta p = A(r,z)\exp[i(m\phi + \sigma t)]$ of the pressure (and any other) perturbation. The corresponding mode numbers are $m$ (axial), $j$ (vertical), and $n$ (radial). One would expect that most mechanisms of excitation (such as turbulence) would produce the highest amplitudes for the lowest modes numbers, where the wavelength of the mode best matches the dominant length scale of the (turbulent, etc.) excitation. In addition, the resulting luminosity fluctuations would tend to increasingly cancel for higher mode numbers.

The g (inertial-gravity)--mode oscillations \citep{per} can be characterized by a restoring force that is dominated by the net gravitational--centrifugal force. The axisymmetric ($m=0$) g--modes are centered at the radius $r_g$ where the radial epicyclic frequency $\kappa$ achieves its maximum. The eigenfrequencies of the modes with only a few nodes in the radial eigenfunction are slightly below $\kappa_0\equiv\kappa(r_g)$.  

The c (corrugation)--modes which have so far been investigated \citep{swo} are non-radial ($m=\pm 1$) vertically incompressible waves near the inner edge of the disk, that precess around the angular momentum of the black hole. They do not exist in counter-rotating ($-1<a<0$) disks. Their fundamental frequency coincides with the Lense-Thirring frequency (produced by the dragging of inertial frames generated by the angular momentum of the black hole) at the outer boundary of their capture zone, in the appropriate slow-rotation limit. 

The p (inertial-pressure)--mode oscillations can be characterized by a restoring force that is dominated by pressure gradients, as for ordinary acoustic modes. The fundamental p--modes exist near the inner and outer edges of the disk, but higher modes can cover a greater area. The frequency of the inner edge fundamental p--mode is comparable to that of the fundamental g and c--modes, and much greater than that of the outer edge p--mode, so only it will be considered below. Details will be presented by \citet{osw}.

Unless otherwise indicated, we take $c=1$, and express all distances in units of $GM/c^2$ and all frequencies in units of $c^3/GM$ (where $M$ is the black hole mass). 

\section{Properties of the Fundamental Modes}

\subsection{g--modes} 

The axisymmetric g--modes are trapped in the region of the disk where the eigenfrequency $|\sigma|\leq\kappa$. The width $\Delta r$ of this region [centered on $r_g(a)$] is proportional to $c_s^{1/2}$. The speed of sound $c_s(r,z)$ is proportional to the disk luminosity $L$ where radiation pressure dominates. This should be valid (for typical luminosities) because $r_g$ is near the radius where the disk temperature peaks. For $0 < a \lesssim 0.95$ and small values of the mode numbers $j$ and $n$, the width \be
\Delta r \approx 3.3(L/L_{Edd})^{1/2} \; ,
\ee 
while $r_g(0)=8$ and $r_g(0.95)\cong 3$ [see \citet{per}, Figure 2].  

The eigenfrequencies of the axisymmetric g--modes with small values of the vertical ($j$) and radial ($n$) mode numbers are 
\be
\sigma_g = (1-\epsilon_{jn})\kappa_0 \; ,\quad  \epsilon_{jn}\approx E_{jn}(a)L/L_{Edd} \; . \label{gfreq}
\ee
The small correction term $\epsilon_{jn}$ depends upon the adiabatic index $\Gamma$ as well as the usual parameters $M$, $a$, the viscosity parameter $\alpha$, and the luminosity $L$. For relevant values and $n\sim j \lesssim 1$, $\epsilon \approx h/r_g$, giving $E_{jn}(0.0)\cong 0.077$ and $E_{jn}(0.9)\cong 0.11$. Thus the frequencies should drop slightly as the luminosity increases (as long as the disk remains thin). In particular,
\be
d\log\sigma_g/d\log L = -E_{jn}(a)L/L_{Edd}  \label{gL}
\ee
when $\epsilon_{jn}\ll 1$.

\subsection{c--modes} 

The fundamental ($|m|=j=1$) c--modes are trapped within $r_i\leq r\leq r_c$. For radial mode number $n=0$, its width can be approximated by the formula
\be
r_c-r_i = K_0a^{-K_1}(1-a)^{K_2} \; , \label{cwidth}
\ee
where $K_0=0.093$, $K_1=0.79$, and $K_2=0.20$ for $M\sim 10M_\sun$ and $L\sim 0.1L_{Edd}$. So the mode has a significant radial extent only for slowly rotating black holes, although this extent increases with increasing radial mode number.

The eigenfrequencies of the fundamental c--modes are also relatively insensitive to the properties of the accretion disk (except for very slowly rotating black holes). They are also related to the basic frequencies mentioned above, in this case by  
\be
\sigma_c = \Omega(r_c)-\Omega_\perp(r_c) \; . \label{cfreq}
\ee
We have found that
\be
d\log\sigma_c/d\log L = -K(a) \; , \label{cL}
\ee
with $K(0.003)=0.51$ and $K(0.1)=0.044$ for $M\sim 10M_\sun$ and $n=0$. The frequencies decrease slowly with increasing values of $n$.

\subsection{p--modes} 

The fundamental ($m=j=0$) p--modes that can produce the high frequencies of interest here also exist near the inner edge of the disk, $r_i\leq r\leq r_p$, where $|\sigma|\geq\kappa$. We take the modified speed of sound 
\be
c_s^*(r) \equiv c_s(r,z=0)/[U^t g_{rr}^{1/2}] \propto (r-r_i)^\mu 
\ee
for $r$ close to $r_i$, where $U^\nu$ is the four-velocity of the fluid and $g_{rr}$ is a metric component. For the standard disk model, $\mu=2/5$. However, magnetic stresses in the gas that is spiraling toward the black hole \citep{hk} may torque the inner edge of the disk enough to produce $\mu=0$. The fraction of the specific angular momentum at $r_i$ that is eventually captured by the black hole is usually denoted by $\beta$. [Thus $\mu(\beta=1)=2/5$ and $\mu(\beta<1)\cong 0$.] For $0\leq\mu<1/2$, we find that
\be 
r_p - r_i = [C_n(\mu)q^{-1}c_s^*(r_p)]^{2/3}  \; , \label{pwidth}
\ee
where $q^2(a)\equiv (d\kappa^2/dr)_{r_i}$ and $C_n\sim(2n+1)[2/(\Gamma+1)]^{1/2}$ for any value of $\mu$. Since typically $r_p-r_i\ll 1$, we expect gas pressure to dominate, giving $\Gamma = 5/3$. For $a\lesssim 0.9$, this equation can approximated by 
\be
r_p - r_i \cong D_n(\beta,L)[a^*_n(\beta,L) - a] \; . 
\ee 
For instance, $D_0(1,0.1L_{Edd}) \cong 0.041$ and $a^*_0(1,0.1L_{Edd}) \cong 1.22$, while $D_0(0.95,0.1L_{Edd}) \cong 0.24$ and $a^*_0(0.95,0.1L_{Edd}) \cong 1.18$. This trapping region size increases (roughly as $n^{2/3}$) for increasing values of the radial mode number $n$ (like the c--modes).  

The eigenfrequencies of the fundamental inner p--modes have a similar dependence on the speed of sound. In terms of the quantities defined above, the basic relation $\sigma^2=\kappa^2(r_p)\cong q^2(r_p-r_i)$ gives
\be
\sigma_p = [C_n(\mu)q^2c_s^*(r_p)]^{1/3} \; , \label{pfreq}
\ee  
so $(r_p-r_i)\sigma_p\cong C_n(\mu)c_s^*(r_p)$. This frequency increases (roughly as $n^{1/3}$) for increasing values of $n$. Since $c_s\propto (L^2/\alpha M)^{1/10}$ near the inner edge,
\be
d\log\sigma_p/d\log L = 1/15 \; . \label{pL}
\ee

\section{Comparison with Observations}

In Table 1 we indicate the radial location and extent of each mode, for two values of the  angular momentum parameter $a$. 
In Figure 1 we plot the fundamental eigenfrequencies of each mode as a function of $a$. All the mode frequencies are essentially proportional to $1/M$.

Since all of the mode frequencies are relatively weak functions of the disk luminosity (and other disk properties such as the viscosity parameter), we shall focus our attention on the two black hole candidates which exhibit more than one `stable' high frequency quasi-periodic oscillation (QPO) in their power spectrum. They are the microquasars GRO J1655-40 \citep{R1655,S1655} and GRS 1915+105 \citep{M1915,S1915}. In Table 2 we present their QPO frequencies $f$, quality factors $Q=f/\Delta f$, and fractional amplitudes. The amplitudes are usually increasing functions of photon energy. 

We can obtain numerical estimates of the dependence of the frequency $f$ on the observed energy flux ${\cal F}$ from equations (\ref{gL}), (\ref{cL}), and (\ref{pL}). For the p--modes, $d\log f/d\log{\cal F} = 0.067$. Note that the frequencies of the g and c--modes decrease with increasing flux. We see that $|d\log f/d\log{\cal F}|\lesssim 0.1$ for them if $L\la L_{Edd}$ (necessary for our thin disk assumption) and $a\ga 0.03$. Only the c--mode can produce a greater variation, for very slowly rotating black holes. Unfortunately, there is little information available on the disk (thermal) flux when each QPO was observed. So all we can do is use these limits to eliminate from consideration any QPOs which violate them. For instance, if we assume that the thermal flux ${\cal F}$ varies by a factor of $<5$, then we require that the frequency $f$ vary by $\la 16$\%. This allows the shift in frequency of about 4\% observed in GRS 1915+105, if the flux was larger when the lower frequency was observed. 

From Table 1, we see that the p--mode probably occupies a smaller region than the other modes. In addition, its oscillations have very little emergent flux to modulate since they are located near the inner edge of the disk. We will therefore focus on the g and c--modes.
Then the pair of frequencies observed in each source allow (at most) two possible choices for $M$ and $a$ (with a spread from unknown properties of the disk), as seen from Figure 1. 

For GRO J1655-40, choosing the lower frequency (295 Hz) to be that of the g--mode [in keeping with our earlier choice \citep{w}] gives $M=5.9\pm 1.0 M_\sun$, $a=0.917\pm 0.024$. This predicted mass is consistent with the value $M=6.3\pm 0.5 M_\sun$ obtained from spectral observations of the companion star \citep{sha,gbo}. (From Figure 1, but for $0.1\leq L/L_{Edd}\leq 1.0$, the corresponding p--mode frequency is then 118---193 Hz for $\beta=0.95$ and 
50.4---87.6 Hz for $\beta=1.00$.) Identifying instead the lower frequency with that of the c--mode gives an unacceptably smaller mass ($\sim 2.5 M_\sun$). 

However, there is an apparent problem with this assignment of the modes. Since the g--mode occupies a much larger region of the disk (Table 1, $a=0.9$), one might expect that the resulting amplitude of the luminosity modulation would be greater than that of the c--mode if they are excited in the same manner. This disagrees with the amplitudes in Table 2. However, since the c--mode is virtually incompressible, it should mainly modulate the coronal photons via reflection from its changing projected area; whereas the g--mode modulates the internal properties of the disk \citep{nw,nwbl}. This matches the fact that the higher frequency QPO has the harder energy dependence \citep{S1655}, since the coronal (power law) photons are more energetic than those emitted directly by the disk. In addition, the viscous damping (or growth) rate of the c--mode should be less than the viscous growth rate of the g--mode \citep{ort}, producing a larger $Q$ value (which agrees with Table 2). Of course, nonlinearities may play an important role. The excitation of higher mode numbers, as well as radiation leakage, can also contribute to the frequency width \citep{nwbl}.    

For GRS 1915+105, our earlier choice \citep{nwbl} of the g--mode for production of the 67 Hz QPO gives $M= 18.2\pm 3.1 M_\sun$, $a=0.701\pm 0.043$. (The corresponding p--mode frequency is then 34.6---56.8 Hz for $\beta=0.95$ and 14.5---25.2 Hz for $\beta=1.00$.) If it is produced by the c--mode, $M=42.4\pm 7.0 M_\sun$, $a=0.926\pm 0.020$. (The corresponding p--mode frequency is then 16.3---26.5 Hz for $\beta=0.95$ and 7.00---12.1 Hz for $\beta=1.00$.) This second choice is supported by the claim of \citet{S1915} that the dependence of the properties of QPOs on their frequency is in the same direction in both sources, as moderately indicated by the data on the quality factors and amplitudes of the QPOs in Table 2. 

Not unexpectedly, the observed power spectra of both sources show no `stable' features at the p--mode frequencies indicated above. An example of another mode of interest is the $m=1$ g--mode, whose frequency is about a factor of 4 greater than that of the $m=0$ (fundamental) g--mode \citep{per}. The large inclination angle of about 70 degrees for both of these microquasars could make this nonaxisymmetric mode (like the c--mode) visible. However, existing power spectra are often too noisy to be able to detect such higher frequencies. 

There is not yet evidence that any of the high-frequency QPOs in other black hole candidates [XTE J1550-564 \citep{R1550,H1550,mil}, XTE J1859+226 \citep{C1859}, and 4U 1630-47 \citep{R1630}] are stable to the extent required if they are produced by these modes. Some are known to vary significantly, while others have only been detected during a single observation.

\acknowledgments

This work was supported by NASA grant NAS 8-39225 to Gravity Probe B.

\begin{deluxetable}{ccccccc}
\tablecolumns{7} 
\tablewidth{0pc} 
\tablecaption{Mode Trapping Regions\tablenotemark{*}} 
\tablehead{ 
\colhead{} & 
\colhead{} &
\multicolumn{2}{c}{Radial location} &
\colhead{} & 
\multicolumn{2}{c}{Radial extent}\\ 
\cline{3-4}
\cline{6-7} \\ 
\colhead{Mode} &
\colhead{$\beta$} &  
\colhead{$a = 0.1$}    &
\colhead{$a = 0.9$}    &
\colhead{}   &
\colhead{$a = 0.1$}    & 
\colhead{$a = 0.9$}      }    
\startdata 
$g$ &  --  & 7.6 & 3.4 &  &  1.0\phantom{00} & 1.0\phantom{00} \\
$c$ & 1.00 & 5.7 & 2.3 &  &  0.56\phantom{0} & 0.064 \\
$p$ & 0.95 & 5.7 & 2.3 &  &  0.26\phantom{0} & 0.067 \\
$p$ & 1.00 & 5.7 & 2.3 &  &  0.046           & 0.013 \\
\enddata
\tablenotetext{*}{For $L\sim 0.1L_{Edd}$, $\alpha\sim 0.1$, and $M\sim 10M_\sun$. Units are $GM/c^2$.} 
\end{deluxetable} 

\begin{deluxetable}{lccc}
\tablecolumns{4} 
\tablewidth{0pc} 
\tablecaption{QPO Properties\tablenotemark{*}} 
\tablehead{ 
\colhead{Source} & 
\colhead{Frequency (Hz)} &
\colhead{$Q$} &  
\colhead{Amplitude (\%)} }    
\startdata 
GRO J1655-40 & 295\phantom{.4} $\pm$ 9\phantom{.5} & $\sim$ \phantom{0}4 & 0.81 $\pm$ 0.11 \\
GRO J1655-40 & 449\phantom{.4} $\pm$ 5\phantom{.5} & $\sim$ 23 & 
 4.8\phantom{0} $\pm$ 0.6\phantom{0} \\
   &   &   &   \\
GRS 1915+105, A01& \phantom{1}66.4 $\pm$ 1.5  & $\sim$ 20 & 6\phantom{.00} $\pm$ 1\phantom{.00} \\
GRS 1915+105, A02& \phantom{1}69.2 $\pm$ 0.2  & $\sim$ 20 & 1.9\phantom{0 $\pm$ 0.11}  \\
GRS 1915+105, A02& \phantom{1}41.5 $\pm$ 0.4 & $\sim$ \phantom{0}8  & 2.4\phantom{0 $\pm$ 0.11} \\ 
\enddata 
\tablenotetext{*}{From \citet{R1655}, \citet{S1655}, \citet{M1915}, and \citet{S1915}. A01 and A02 refer to different epochs of observation.}
\end{deluxetable} 

\begin{figure}
\figurenum{1}
\epsscale{0.98}
\plotone{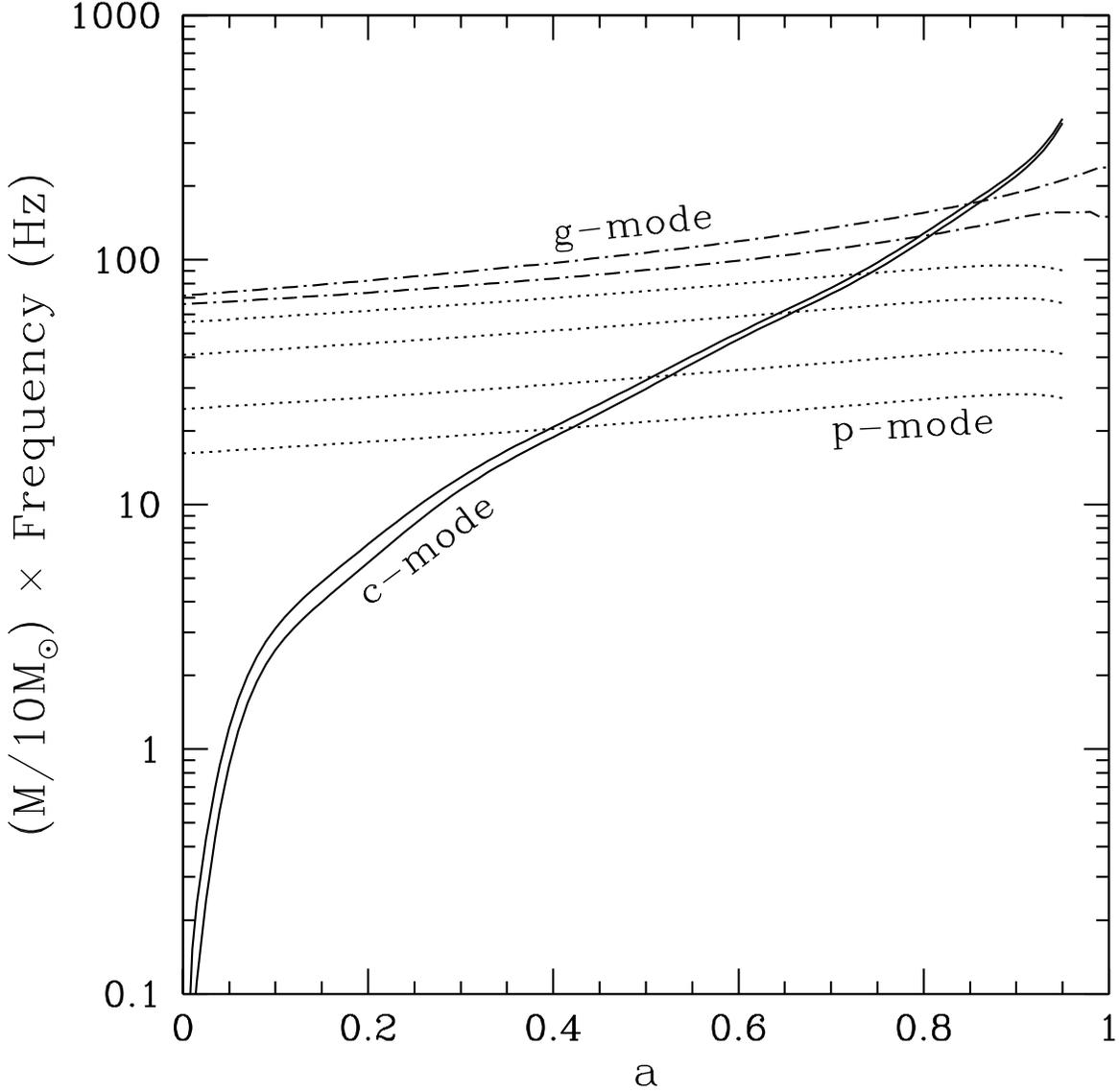}
\caption{The dependence of the fundamental frequencies of the three diskoseismic modes on the angular momentum of the black hole.  The spread of each corresponds to the range $0.01\leq L/L_{Edd}\leq 1.0$, and is relatively insensitive to the choices $M/M_\sun\sim 10$ and $\alpha\sim 0.1$. The upper band for the p--mode corresponds to $\mu=0$ with $\beta= 0.95$, and the lower band to $\mu=2/5$ (no torque, $\beta=1.00$).}
\end{figure}

\end{document}